\documentclass[conference]{IEEEtran}

\IEEEoverridecommandlockouts
\usepackage{color}
\usepackage{graphicx}
\usepackage{xcolor}
\usepackage{url}
\usepackage[colorlinks=true,linkcolor=black,urlcolor=black,citecolor=blue]{hyperref}
\usepackage{cite}
\usepackage[english]{babel}
\usepackage[utf8]{inputenc}
\usepackage{amsmath,amssymb}
\usepackage{algorithm}
\usepackage{algpseudocode}
\usepackage{multirow}
\usepackage{svg}
\usepackage{graphicx}
\usepackage{caption}
\usepackage{subcaption}
\usepackage{dblfloatfix}

\usepackage{blindtext}

\definecolor{darkblue}{rgb}{0, 0, 0.5}

\usepackage{todonotes}

\begin{document}

\title{A Novel Non-Terrestrial Networks Architecture:\\
All Optical LEO Constellations with \\High-Altitude Ground Stations}

\author{
\IEEEauthorblockN{
Pablo G. Madoery\IEEEauthorrefmark{1}\IEEEauthorrefmark{3},
Juan A. Fraire\IEEEauthorrefmark{2}\IEEEauthorrefmark{3},
Jorge M. Finochietto\IEEEauthorrefmark{3},
Halim Yanikomeroglu\IEEEauthorrefmark{1},
Gunes Karabulut Kurt\IEEEauthorrefmark{1}\IEEEauthorrefmark{4}
}

\IEEEauthorblockA{
\IEEEauthorrefmark{1} Non-Terrestrial Networks Lab, Department of Systems and Computer Engineering, Carleton University, Ottawa, Canada\\
\IEEEauthorrefmark{2}Inria, INSA Lyon, CITI, UR3720, 69621 Villeurbanne, France\\
\IEEEauthorrefmark{3}CONICET - Universidad Nacional de Cordoba, Cordoba, Argentina\\
\IEEEauthorrefmark{4}Poly-Grames Research Center, Department of Electrical Engineering, Polytechnique Montréal, Montréal, Canada
}}

\maketitle

\begin{abstract}
% Context
The emergence of low Earth orbit (LEO) satellite mega-constellations is dynamically transforming the space sector. 
% Problems  
While free-space optical (FSO) links efficiently facilitate inter-satellite data forwarding, they suffer from atmospheric/weather conditions in the space-to-ground link. 
% Approach
This study delves into utilizing high-altitude platform stations (HAPS) as elevated relay stations strategically positioned above terrestrial ground stations. 
We introduce the concept of high-altitude ground stations (HAGS), an innovative approach to enabling the development of all optical LEO satellite constellations.
% Contribution 1)
The first contribution is an analysis of the HAGS-based network architecture where the LEO spacecraft only hosts FSO transceivers. 
% Contribution 2)
Secondly, we execute an extensive simulation campaign to determine the gain of HAGS, including a new equivalency model with the traditional ground station approach. 
% Contribution 3)
Finally, we examine the research challenges of implementing HAGS-based, all optical LEO mega-constellations.
\end{abstract}


%%%%%%%%%%%%%%%%%%%%%
% Introduction
%%%%%%%%%%%%%%%%%%%%%
\section{Introduction}
\label{sec_introduction}

% HAPS Motivation
Integrating high-altitude platform stations (HAPS) ascending beyond the stratosphere has significantly invigorated the networking field~\cite{kurt2021vision}. 
% HAPS in Mobile (HIBS)
HAPS has been notably effective in their versatile applications as cellular base stations, specifically as international mobile telecommunications base stations (HIBS)~\cite{EulerHibs}. 
% HAPS in Space (HAGS)
Building on this concept, this paper ventures into the space networking and communication domain, proposing using HAPS as elevated ground stations (GS), or HAGS. 

\begin{figure*}[]
  \centering
  \includegraphics[width=1\textwidth]{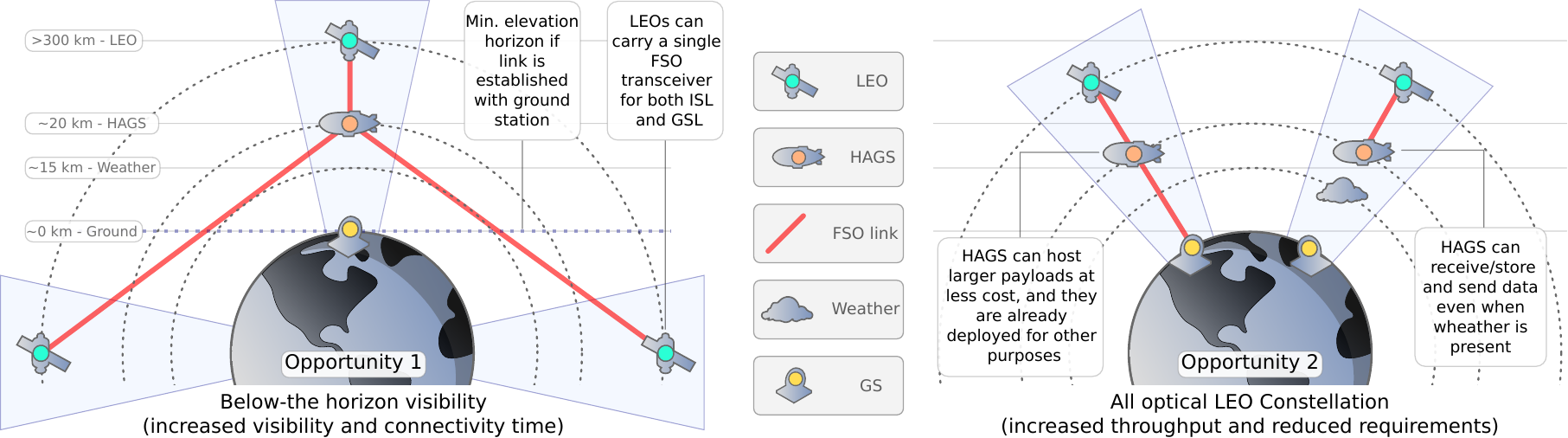}
  \caption{Opportunities of HAGS: enhanced visibility time and weather impairment buffer.}
  \label{fig:benefits}
\end{figure*}

% Goal: FSO Links
One of HAGS's most groundbreaking opportunities is enabling all free-space optical (FSO) LEO constellations.
% ISL and SGL
This development encompasses not only inter-satellite links (ISLs) but also, crucially, ground-satellite (GSL) feeder links, which have traditionally been hindered by weather-related impairments.
% FSO Only transponder, no RF
In a HAGS-based setup, satellites can use the same FSO interface for ISL and HAGS links, eliminating the need for backup or hybrid RF systems.
% Benefits
%This capability is twofold: firstly, it mitigates weather-related impairments; secondly, it extends visibility time due to the higher altitude of HAGS (see Fig.~\ref{fig:benefits}).
This capability has two aspects: mitigating weather-related impairments and extending visibility time due to HAGS' higher altitude (see Fig.~\ref{fig:benefits}).

% Network: Store-Carry and Forward
HAGS serve a dual function, acting as a relay and a data buffer. 
HAGS facilitates rate and weather decoupling, ensuring data delivery even when satellites are out of reach from the GS. 
This feature is particularly advantageous during periods of limited connectivity imposed by orbital dynamics, allowing for the maximization of LEO-to-HAGS throughput. 
This aspect is particularly relevant in Earth observation missions where data can flow in a store-carry-and-forward fashion.
% Other features of HAGS
The deployment of HAGS also broadens the scope of satellite communication systems. 
By potentially serving as HIBS, HAGS reduces the barrier to investment and expands operational capabilities. 
Moreover, HAGS offers the capability to download and potentially process bulk data directly on the HAGS. 
This approach accelerates data availability and leverages HAGS as mission data centers from which data can be accessed and analyzed.

% Contribution
This paper contributes significantly to the advancement of HAGS towards all optical constellations.
% Architecture
Firstly, we provide a detailed description of the HAGS network architecture, laying the foundation for understanding its operational framework and potential applications.
% Performance Evaluation
Secondly, we conduct a comprehensive performance evaluation of HAGS, employing model-based simulations to compare its efficacy with traditional GS. This comparison quantifies the benefits of HAGS and enables a GS-to-HAGS equivalency model, offering valuable insights for decision-makers in the field. 
% Challenges
Lastly, we identify and discuss the open challenges within the HAGS network, highlighting areas for future research and development. 

% Organization
The rest of the paper is organized as follows.
Section~\ref{sec_background} offers a detailed overview of HAPS and FSO satellite communication. 
We delve into the system model in Section~\ref{sec_model}. Section~\ref{sec_evaluation} presents our evaluations and a model-based equivalency analysis. 
Section~\ref{sec_challenges} highlights and discusses the prevailing challenges in HAGS. 
Finally, Section~\ref{sec_conclusion} encapsulates our research's key findings and conclusions.

%%%%%%%%%%%%%%%%%%%%%
% Background
%%%%%%%%%%%%%%%%%%%%%
\section{Background}
\label{sec_background}

% Use case 1: NTN
Non-terrestrial networks (NTN) will be a fundamental segment of future 5G/6G systems~\cite{kota20216g}, which require efficient bidirectional data transport mechanisms via efficient feeder links to the GS.
% Use case 2: EO
Likewise, a disruptive shift toward constellations of small spacecraft with increasing Earth observation data volume~\cite{marcuccio2019smaller} further stresses data rate requirements.
Whether for NTN or Earth observation, RF and FSO feeder links, to and from the GS, are pushed to the limit.

% RF solutions
\paragraph*{RF and FSO}
Although the predominant approach for space-to-ground data transfer has traditionally centered around RF bands such as Ku and Ka, \cite{zedini2020performance} highlight the limitations in capacity have largely reached a saturation point.
Efforts are underway to investigate higher bands, such as Q/V/W, which offer broader spectrum availability, to enhance the capacity of very high throughput satellite (VHTS) feeder links. 
However, the propagation impairments experienced at tropospheric altitudes (ranging from 8 to 15 km) necessitate implementing spatial diversity techniques across multiple ground gateways.
% FSO solutions
%\paragraph*{FSO}
In this context, FSO links have garnered considerable attention as pivotal for HAPS~\cite{alzenad2018fso} and LEO satellites~\cite{toyoshima2022applicability}. % ~\cite{chaudhry2020free,chaudhry2021laser,toyoshima2022applicability}.
% FSO for GSL
However, when utilized in space-to-ground scenarios, FSO encounters significant challenges associated with pointing, beam wander~\cite{zedini2020performance}, and weather impairments~\cite{liang2022link}, requiring the employment of sophisticated acquisition, pointing, and tracking (APT)~\cite{kaushal2016optical}, signal processing~\cite{paillier2020space} and link diversity techniques~\cite{gong2019network}.
% Hybrid FSO
As a result, when considered GSLs, FSO systems are typically framed in complex hybrid FSO/RF systems where backup RF links are used during inclement weather conditions~\cite{kazemi2014outage}. 

% HAPS
\paragraph*{HAPS}
By utilizing platforms positioned in the stratosphere at approximately 20 km altitude~\cite{alam2021high}, HAPS can enhance communication coverage over large regions~\cite{kurt2021vision}.
While HAPS are commonly associated with free-flying configurations, alternative concepts involving tethered balloons or blimps are also being contemplated to take advantage of a constant supply of power and data through the tether. 
Nevertheless, the presence of the tether imposes altitude restrictions, typically up to 5 km~\cite{belmekki2022unleashing}.
% HAPS in Networks
HAPS have been widely studied as part of a space-air-ground integrated network, surveyed in~\cite{liu2018space}.
Implementing optical terminals in HAPS to link with LEO satellites dates back to 2007 when authors in~\cite{giggenbach2007optical} introduced the stratospheric optical relay stations. 
Since then, several contributions have been made at the system, application, and link (optical only and hybrid) levels.

% System level
\paragraph*{System and Applications}
Multiple works aimed at linking HAPS and satellites. \cite{agnew2012edrs} state that the European Data Relay System (EDRS) is already providing HAPS to satellite data relay services.
% MILP
On the research front, cooperative HAPS and LEO were assessed using a time-evolving graph, which is used for resource allocation via mixed integer linear and non-linear programming in~\cite{jia2021toward}.
% HAPS-LEO spectrum
Authors in~\cite{li2020hierarchical} also propose a dynamic resource optimization method for shared spectrum utilization between LEO and HAPS. 
% HAPS trajectory
%Also, HAPS trajectory optimization using deep %reinforcement learning in a HAPS-satellite %integrated system was proposed in %\cite{lee2020integrating}.
% Collaboration
Collaborative HAPS-satellite task offloading in networks with connected HAPS and LEO satellites was studied in~\cite{zhang2020satellite}.
\cite{elmahallawy2022fedhap} introduced a collaborative federated learning framework between HAPS and LEO satellites.
% Security
In the security domain, authors in~\cite{yahia2021use} proposed optically-linked satellites and HAPS to increase secrecy in satellite networks. 
% QKD
% Also, quantum key distribution (QKD) systems leveraging HAPS and FSO were studied in~\cite{vu2020design}.

% % Terminal
% \paragraph*{Terminal Design}
% Design guidelines for a HAPS-to-LEO transceiver are provided in~\cite{antonini2006feasibility}. Optical window frequencies with minimal absorption and scattering at 20 to 100 km altitudes are investigated. 
% Store-carry-and-forward relay via RF links from HAPS to the ground is mentioned but not evaluated.
% A subsequent publication succinctly outlined the prerequisites for spatial acquisition and environmental attributes~\cite{perlot2008system}.

% Optical Link
\paragraph*{Link Models}
Authors in~\cite{fidler2010optical} discussed the challenges in APT and the influence of the atmosphere, background light, and flight qualification requirements in HAPS-to-LEO FSO links.
A two-hop (satellite-to-HAPS-to-ground) assessment was made in~\cite{vu2018performance}. 
The focus is on transparent and regenerative all-optical relay solutions, but no storage or buffering is considered in the HAPS.
% HAPS FSO to ground
An analytical channel model and link design optimization for ground-to-HAPS free-space optical communication networks were presented in~\cite{safi2020analytical}.
The point-to-point optical and radio link budget between HAPS, UAVs, and satellites on different layers is studied in~\cite{ata2022haps}.
The paper in~\cite{erdogan2021site} finds that the exponentiated Weibull (fading best fits different aperture sizes in most weather conditions.

\begin{figure*}[]
  \centering
  \includegraphics[width=\textwidth]{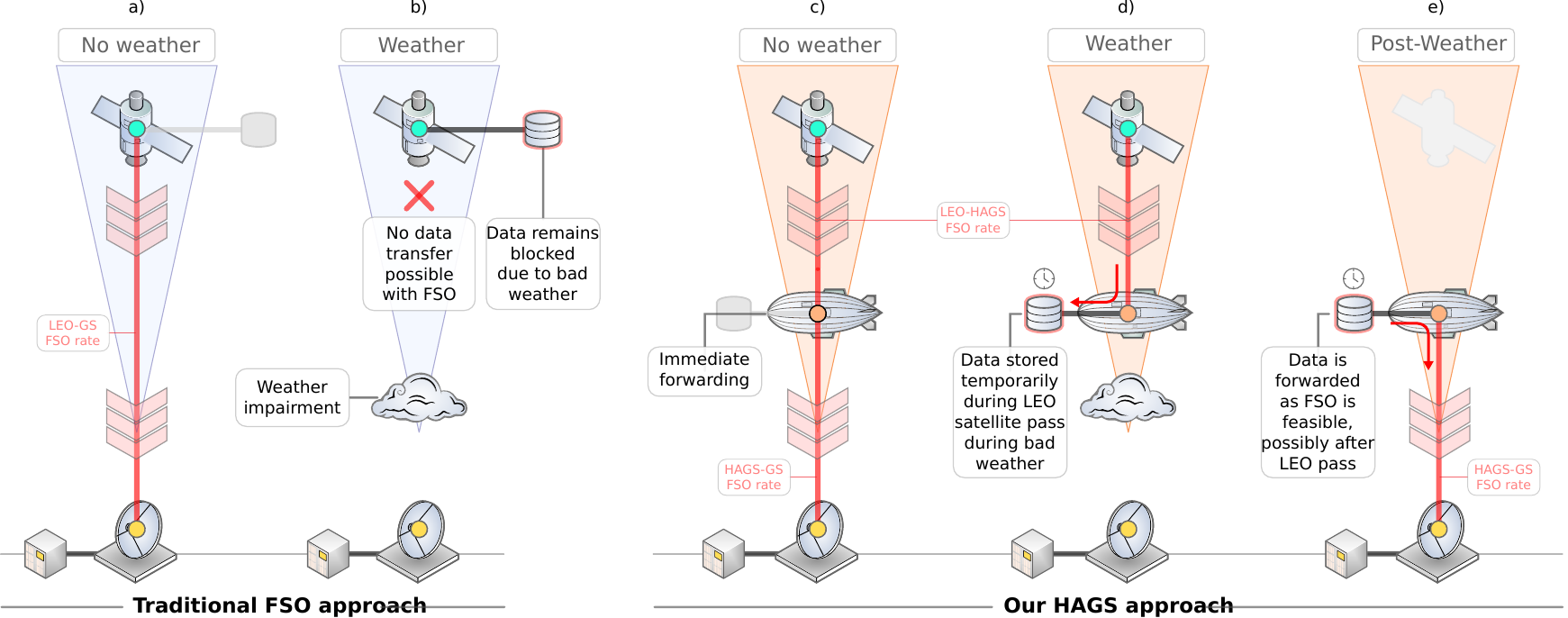}
  \caption{Downlink data handling and operations in traditional FSO and our HAGS approach.}
  \label{fig_states}
\end{figure*}

% Hybrid
\paragraph*{Hybrid FSO/RF}
Connecting HAPS, LEO, and GS with hybrid FSO/RF is discussed in~\cite{swaminathan2020performance}.
The authors later presented an analysis for a dual-hop uplink in~\cite{swaminathan2021haps}, confirming hybrid FSO/RF can improve the performance of HAPS-to-LEO relay links.
\cite{shah2021adaptive} considered Hybrid FSO/RF for uplink and downlink with a focus on symbol error rates.
While these papers assumed a gamma–gamma distribution for FSO communication, a follow-up paper \cite{yahia2022haps} considers the exponential Weibull fading, which provides a better fit for terminals with larger apertures. 

%Despite the growing interest in the research community around integrating HAPS and LEO satellites with FSO links, a critical aspect has been consistently overlooked: the potential of HAGS to enable all optical LEO constellations. 
%This oversight is significant, as the ability of LEO nodes to utilize optical links for both ISLs and GSLs could profoundly impact the sector. Yet, it remains largely unexplored in existing research. The remainder of this paper seeks to cover this gap.
Despite increasing research interest in integrating HAPS and LEO satellites with FSO links, one critical aspect has been consistently overlooked: the potential of HAGS to enable all-optical LEO constellations. This oversight is significant, as it could profoundly impact the sector by allowing LEOs to utilize optical links for both ISLs and GSLs. The remainder of this paper aims to address this gap.

%%%%%%%%%%%%%%%%%%%%%
% Model
%%%%%%%%%%%%%%%%%%%%%
\section{System Model}
\label{sec_model}

% \subsection{Network Model}

Fig.~\ref{fig_states} illustrates the traditional and HAGS-based models that are the focus of our evaluation in this paper. Each LEO satellite within the constellation is equipped with an FSO terminal in the depicted models. 
This terminal enables the satellite to transmit data directly to a GS or HAGS.

% Traditional
\paragraph*{Traditional Model}
When considering a traditional FSO scheme \cite{Airbus2018}, LEO satellites reach one of the multiple GS via episodic contacts. If there is line-of-sight between a LEO and a GS, the LEO can use the contact to download data at the highest possible data rate (see Fig.~\ref{fig_states}-a).  
Suppose optical communication is not possible due to cloud coverage (see Fig.~\ref{fig_states}-b). In that case, data will remain in the LEO, which will take advantage of subsequent contact with another GS.

% HAGS
\paragraph*{HAGS Model}
Each HAGS is positioned 20 km above a corresponding GS, strategically placed above cloud level (see Fig.~\ref{fig_states}-c). 
This positioning assumes that LEO-HAGS contacts remain consistently available, provided a line-of-sight exists. 
However, the connectivity between each HAGS and GS depends on cloud cover. 
In scenarios where contact is temporarily unavailable, HAGS can store data in a buffer (see Fig.~\ref{fig_states}-d) and forward it to the GS once the link is re-established (see Fig.~\ref{fig_states}-e). 
This store-carry-and-forward capability, a cornerstone of the delay-tolerant networking (DTN) architecture~\cite{RFC4838}, is effectively utilized in tandem with advanced routing algorithms like contact graph routing (CGR)~\cite{Fraire2021Routing}. 
CGR dynamically computes forwarding policies based on a pre-calculated and distributed contact plan, ensuring efficient data transmission across the network nodes.

% \subsection{Cloud Cover Model}
% \label{sec_model_cloud}
\paragraph*{Cloud Cover Model}
Cloud cover significantly impacts the stability of FSO links, with clouds acting as a primary obstruction. The unpredictability of cloud cover, its varying duration, and the likelihood of clear skies can be modeled using an exponential (Poisson) distribution. 
It delineates two fundamental outcomes: a 'blocked' state during cloudy conditions and an 'unblocked' state under clear skies. 
Characterized as a \textit{memoryless} distribution, this weather model effectively captures the sporadic nature of cloud cover without any dependency on past events.
The model takes a single known average rate parameter for cloud cover and can be described by
$C(t)= \int^{t}\mu e^{-\mu \tau} = 1 - e^{-\mu \tau}$.
In this equation, $h(t)=\mu$ represents the \textit{cloud cover rate}, and $\tau$ is the \textit{time}. The cloud cover rate simplifies to the constant $\mu$ for any time. 
The \textit{mean time to cloud cover} (\textbf{TCC}) is $TCC=1/\mu$. 
Additionally, we generate the \textit{mean time to clear sky} (\textbf{TCS}) using the exponential model. 
TCC and TCS are the core input parameters for the cloud cover model.
Like a contact plan, a \textit{weather plan} can be generated using TCC, TCS, and the weather model. This plan includes all predicted cloud cover events.

%%%%%%%%%%%%%%%%%%%%%
% Evaluation
%%%%%%%%%%%%%%%%%%%%%
\section{Evaluation and Analysis}
\label{sec_evaluation}

% Simulator tool
For the comparative analysis of the models outlined in Section~\ref{sec_model}, we extended DtnSim, a discrete event-driven simulator based on the Omnet++ framework. 
Originally introduced in~\cite{Fraire:2017:DtnSim} and accessible via a public repository\footnote{DtnSim is publicly available at:~\url{https://gitlab.inria.fr/jfraire/dtnsim.git}}, DtnSim models each node (be it a LEO, a HAGS, or a GS) as an Omnet module. 
These modules have sub-modules for data generation, routing, and forwarding. 
We implemented the cloud cover module to study the weather conditions in line with the TCC and TCS parameters discussed in Section \ref{sec_model}.

% Topology
\paragraph*{Invariants}
Utilizing Systems Tool Kit (STK), we simulated a week-long interaction between the Iridium constellation and the KSAT's GS network. 
Additionally, we modeled interactions between the same LEO satellite and a series of HAGS, each situated 20 km above its corresponding GS. 
The satellite must evacuate a traffic load of 50 files, each 100 GB. 
All optical links (LEO-GS, LEO-HAGS, and HAGS-GS) operate at a data rate of 8~Gbps. 
These links are assumed to be ideal, with interruptions occurring solely due to cloud cover between the LEO and GS or between the HAGS and GS.

% Control variables
\paragraph*{Variables}
To facilitate a comprehensive comparison, the traditional architecture varies the number of GS (1, 2, 5, and 10), while the HAGS-inclusive architecture adjusts the number of HAGS (1 to 5, each over its respective GS). The TCC is altered between 0.1 and 40 hours, and the TCS ranges from 5 to 25 hours. Each configuration undergoes 100 iterations to ensure statistical significance, with performance metrics averaged and 95\% confidence intervals included.

%% Performance metric
\paragraph*{Metrics}
The evaluation of performance is based on three key metrics: \textit{a}) Delivery Ratio (\textbf{DR}), which is determined by the percentage of generated files that successfully reach their intended destination, \textit{b}) Delivery Delay (\textbf{DD}), calculated by subtracting the file's generation time from its arrival time, and
mean and maximum values of \textit{c}) Buffer Occupation (\textbf{BO}).

\subsection{Results Analysis}

\begin{figure}[!b]
\begin{subfigure}{.5\textwidth}
  \centering
\includegraphics[width=0.95\linewidth]{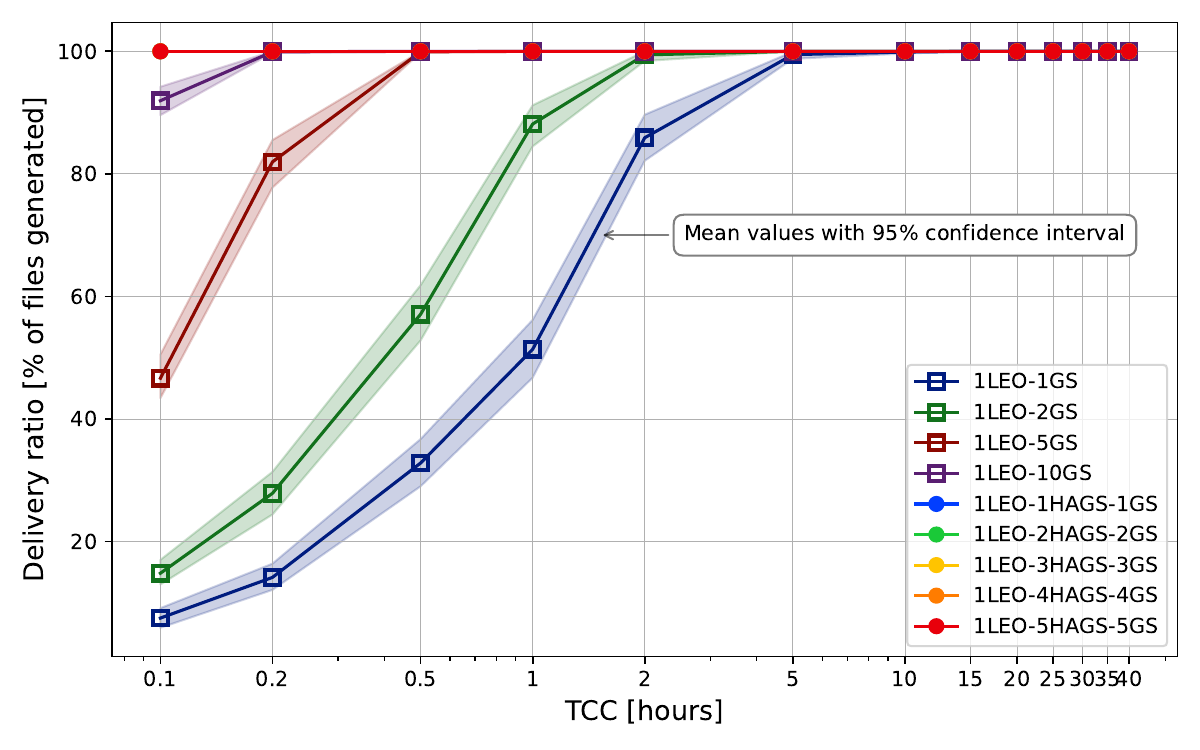} 
  \caption{Delivery ratio with TCS=5 hours.}
  \label{fig:delivery_ratio_5}
\end{subfigure}
\begin{subfigure}{.5\textwidth}
  \centering
  \includegraphics[width=0.95\linewidth]{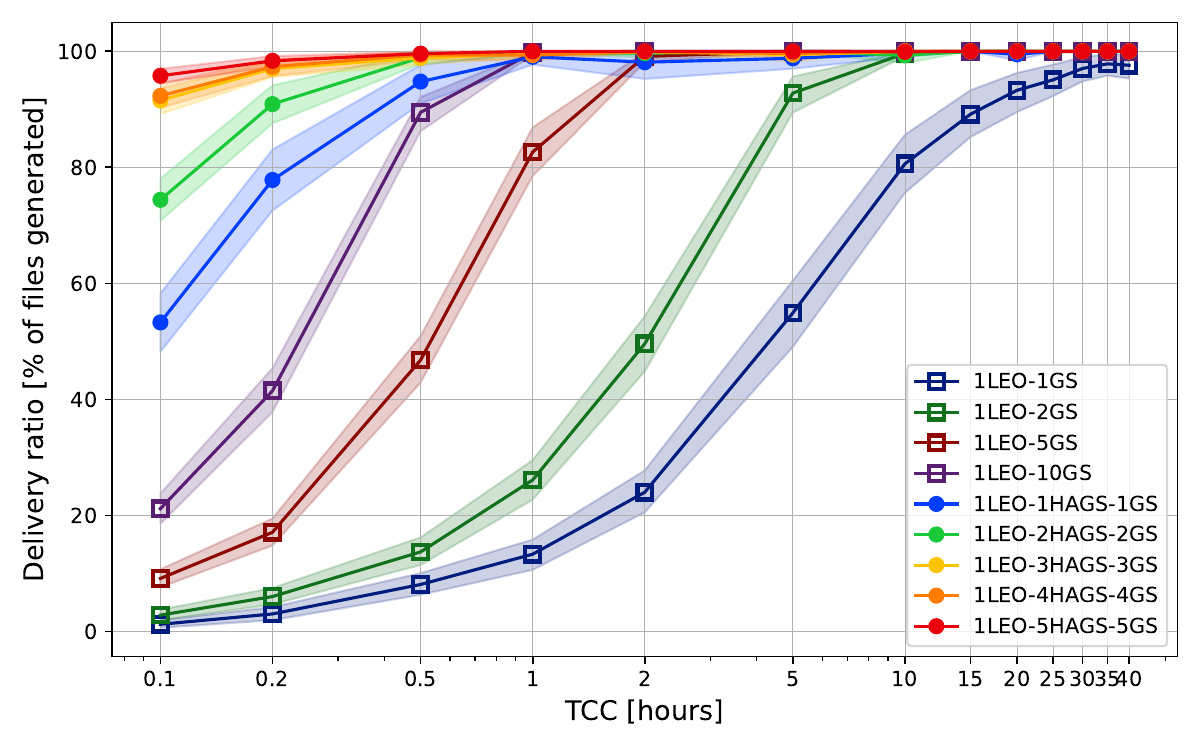} 
  \caption{Delivery ratio with TCS=25 hours.}
  \label{fig:delivery_ratio_25}
\end{subfigure}
\caption{Delivery delay for different TCS and TCC values.}
\label{fig:delivery_ratio}
\end{figure}

\begin{figure}[!b]
\begin{subfigure}{.5\textwidth}
  \centering
  \includegraphics[width=0.95\linewidth]{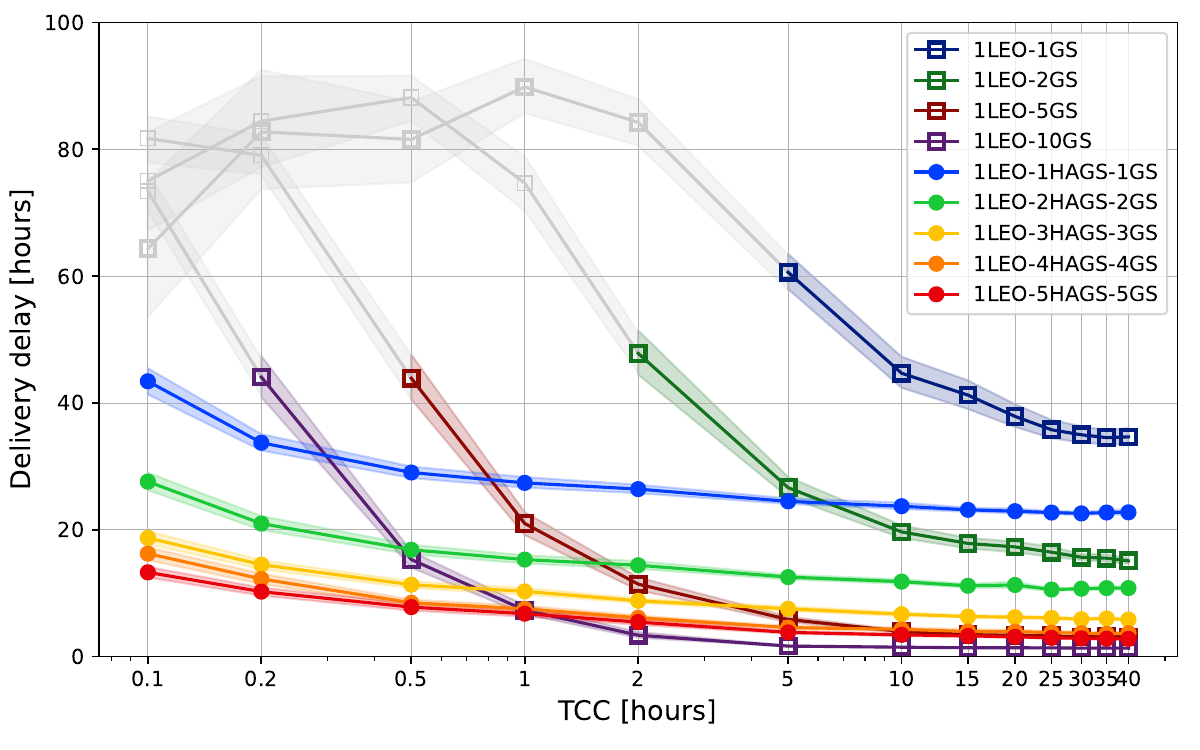} 
  \caption{Delivery delay with TCS=5 hours.}
  \label{fig:delivery_delay_5}
\end{subfigure}
\begin{subfigure}{.5\textwidth}
  \centering
  \includegraphics[width=0.95\linewidth]{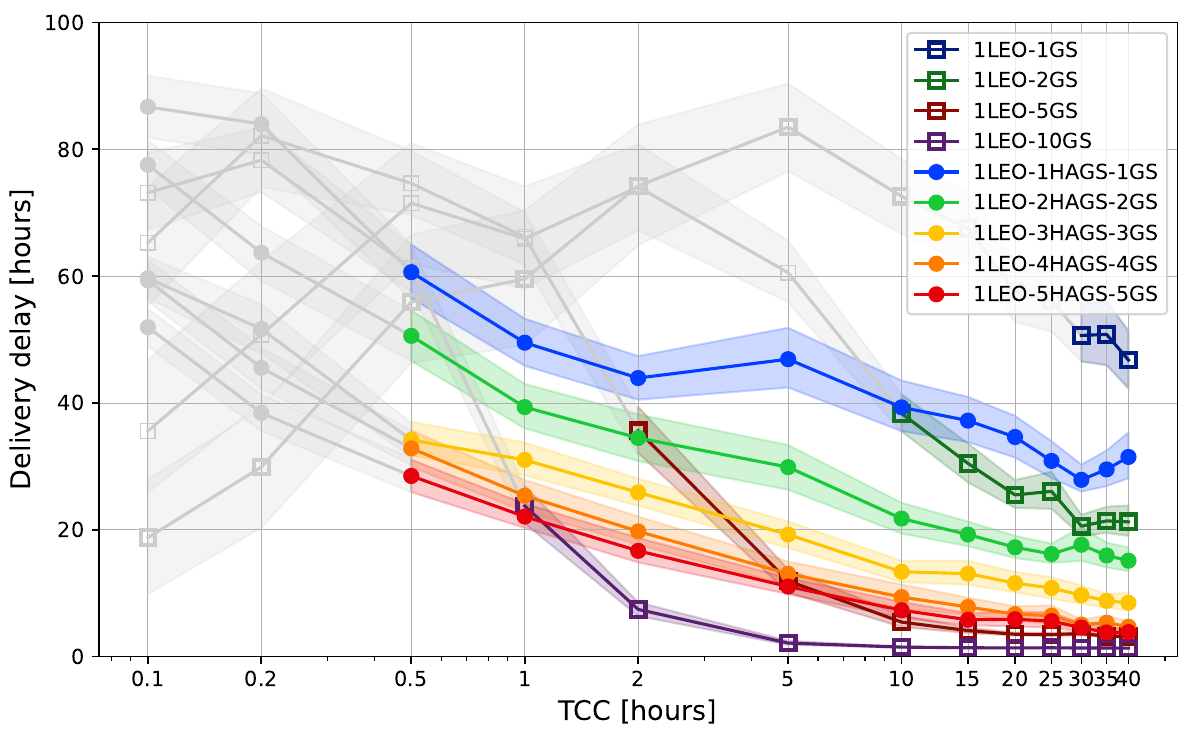} 
  \caption{Delivery delay with TCS=25 hours.}
  \label{fig:delivery_delay_25}
\end{subfigure}
\caption{Delivery delay for different TCS and TCC values.}
\label{fig:delivery_delay}
\end{figure}

\paragraph{Delivery Ratio}
% General
Fig.~\ref{fig:delivery_ratio} illustrates the DR achieved under varying TCC and TCS values. Generally, a lower TCC (indicating a higher cloud cover rate) results in fewer files reaching their destination. 
Conversely, higher TCC values correlate with a 100\% delivery ratio across all schemes. 

% HAGS DR
As depicted in Fig.~\ref{fig:delivery_ratio}-a), HAGS schemes maintain a complete DR even at low TCC values, provided the TCS is fixed at 5 hours. 
This efficiency is attributed to the LEO-HAGS link's immunity to weather disruptions, allowing HAGS to store files until the cloud-affected HAGS-GS link becomes available. 
Thus, LEO satellites can consistently rely on the LEO-HAGS link for transferring data to the ground, irrespective of temporary adverse conditions on the HAGS-GS link.

% GS DR
In contrast, schemes relying solely on GS lack this benefit. Cloud cover impeding LEO-GS communication halts data download, forcing the LEO to await another communication window with a different GS. 
This limitation is more pronounced with fewer GS. Remarkably, even with 10 GS, these schemes underperform compared to HAGS in high cloud cover scenarios (TCC=0.1 hour), underscoring HAGS's superior reliability in adverse weather conditions.

% TCS 25 hrs
Fig.~\ref{fig:delivery_ratio}-b) demonstrates that while all schemes experience reduced performance with longer TCS, those relying solely on GS suffer more significantly than those incorporating HAGS. 
For instance, in the 1LEO-1GS scheme with a constant TCC of 5 hours, the delivery ratio dramatically falls from 100\% to below 60\% as TCS extends from 5 to 25 hours. 
In contrast, the 1LEO-1HAGS-1GS scheme maintains a steady 100\% DR across both TCS values, underscoring the gains of incorporating HAGS into the communication architecture.

\paragraph{Delivery Delay}
% General
Fig.~\ref{fig:delivery_delay} presents the DD per file across varying TCC and TCS values. 
For an accurate comparison, sections of the curves representing DR below 100\% (see Fig. \ref{fig:delivery_ratio}) are greyed out. 
This approach ensures that only schemes delivering an equal number of files are compared. 
The figure reveals that a decrease in TCC generally increases DD for all schemes. 
Notably, schemes incorporating HAGS consistently exhibit shorter DD than those using only GS, particularly at higher cloud rates (lower TCC values).

% Short TCS = 5 hours
Examining Fig.~\ref{fig:delivery_delay}-a), we observe that at a TCC of 0.5 hours, the performance of the 2 HAGS scheme closely matches that of the 10 GS scheme. 
This suggests that a higher number of GS is necessary to match the efficiency of a smaller number of HAGS, particularly during periods of high cloud cover. 
As TCC values increase, this disparity diminishes. 
For instance, with a TCC of 2 hours, the performance of 2 HAGS aligns with that of 5 GS, and at a TCC of 5 hours, a single HAGS is roughly equivalent to 2 GS in terms of delivery efficiency.

% Short TCS = 5 hours
Fig.~\ref{fig:delivery_delay}-b), focusing on a TCS of 25 hours, reveals interesting dynamics in DD. 
When setting TCC to 5 hours and extending TCS, the performance gap between the 4 HAGS scheme and the 10 GS scheme widens, increasingly favoring the GS-based approach. 
This shift occurs because leveraging geographical diversity by routing traffic through multiple GS becomes more advantageous in scenarios with less frequent cloud cover and prolonged TCS. 
This strategy allows quicker offloading of data compared to waiting for the availability of a HAGS-GS link. 
Consequently, the benefits of using HAGS over multiple GS become more pronounced at lower TCC and TCS values.

\paragraph{Buffer Occupation} Measurement of buffer occupancy when using one HAGS reveals that for small values of TCC, the maximum occupancy can reach up to 60\% of the total traffic generated, while the average occupancy reaches up to 10\%. These values decrease to maximums of 15\% and averages of 1\% when TCC is greater than 20 hours.  

\subsection{Equivalency Analysis between GS and HAGS}

This section analyzes the equivalency between GS and HAGS. 
We aim to ascertain the requisite number of GS and HAGS to achieve comparable DD and DR.
To achieve this, we initially identify the intersection points of all curves in Figs.~\ref{fig:delivery_delay} and~\ref{fig:delivery_ratio} across all simulation iterations. 
This process yields a dataset comprising nearly 20,000 equivalency points, where configurations of GS and HAGS demonstrate similar DR and DD. 
We employ this dataset to train a Support Vector Machine (SVM) model, utilizing a Radial Basis Function (RBF) kernel. 
The RBF kernel is chosen for its ability to measure similarity based on the spatial proximity of data points.

\begin{figure}
  \centering
  \includegraphics[width=0.9\linewidth]{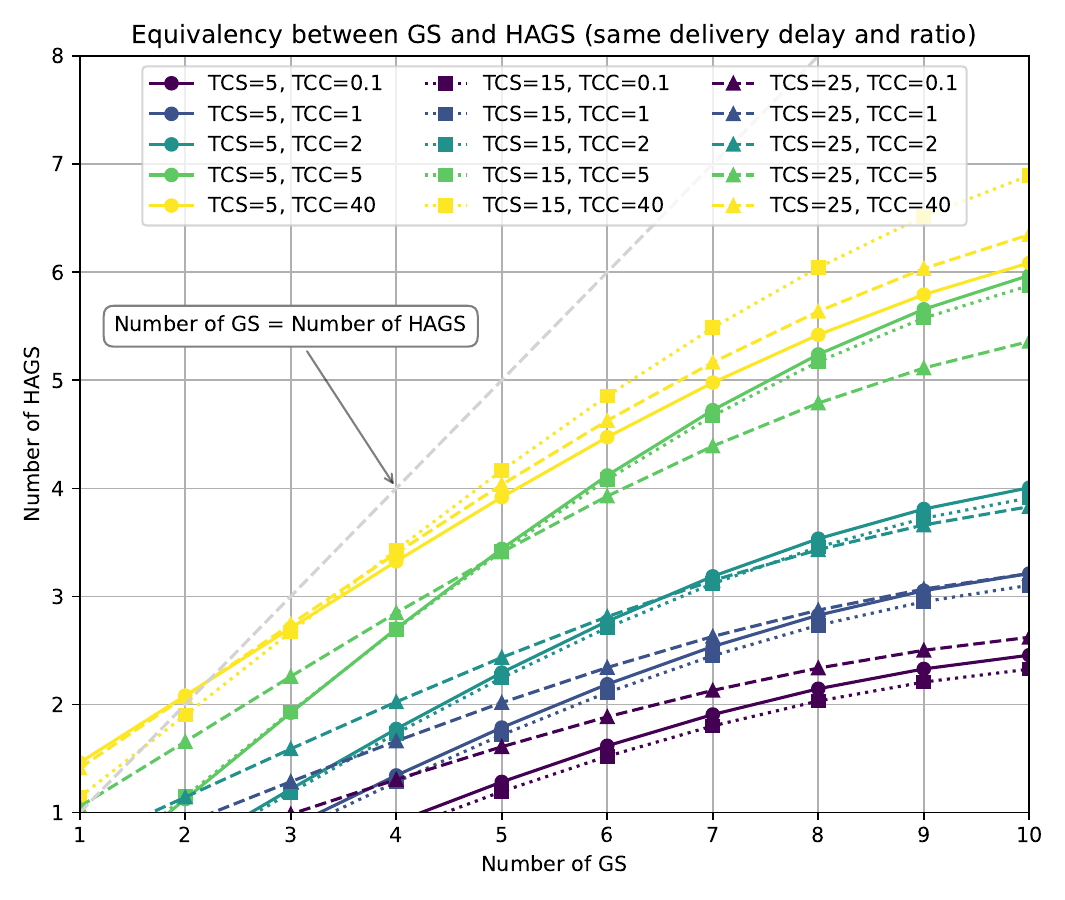}
  \caption{Equivalency Analysis between GS and HAGS.}
  \label{fig:equivalency}
  \vspace{-3 mm}
\end{figure}

The generalized equivalency results, as deduced by this model, are depicted in Fig.~\ref{fig:equivalency}. 
The illustration reveals that under most combinations of TCC and TCS, fewer HAGS are required to match the performance of a larger number of GS. 
For instance, it is observed that 2 HAGS can deliver performance on par with 8 GS for TCS ranging between 5 and 15 hours and a TCC of 0.1. 
Notably, as weather conditions ameliorate (characterized by higher TCC and shorter TCS), the advantage of using HAGS diminishes. 
Yet, they remain a more efficient alternative than a larger array of GS, particularly for connecting all optical LEO constellations.

%%%%%%%%%%%%%%%%%%%%%
% Challenges
%%%%%%%%%%%%%%%%%%%%%
\section{Open Challenges}
\label{sec_challenges}

%Even though promising, implementing HAGS presents several key challenges that need addressing to harness their potential fully. 
While promising, implementing HAGS poses significant challenges that must be addressed to fully leverage their potential.
% DTN
Firstly, HAGS necessitates sophisticated store-carry-and-forward end-to-end mechanisms integral to DTN, especially when considering ISLs. 
% Congestion Aware
This includes developing congestion-aware routing strategies to efficiently direct data via ISL to less occupied buffers among available HAGS and effective buffer status updating methods. 
% QoS and Priority
Additionally, there is a critical need for prioritizing the handling of real-time telemetry and commands, ensuring their immediate transmission takes precedence over regular payload or user data.
% Topology design
Another significant challenge lies in the design of HAGS network topology, especially for large-scale mega-constellations. %As the number of satellites passing by often exceeds the number of interfaces on the HAGS, it becomes crucial to devise mechanisms that determine optimal satellite connections, ensuring efficient network management and data flow.
With more satellites than interfaces on the HAGS, it's crucial to devise mechanisms for optimal satellite connections to ensure efficient data flow.
Moreover, the physical layer of HAGS-to-LEO links poses challenges, such as pointing, acquisition and tracking of a not completely stationary HAGS. 
%These links are particularly demanding due to higher angular rates, which necessitate advanced pointing acquisition and tracking procedures and more robust Doppler compensation techniques. 
%
Addressing these challenges is essential for the reliable and effective operation of HAGS in the dynamic environment of LEO satellite constellations.

%%%%%%%%%%%%%%%%%%%%%
% Conclusion
%%%%%%%%%%%%%%%%%%%%%
\section{Conclusion}
\label{sec_conclusion}

This paper proposed high-altitude ground stations (HAGS) as a transformative element towards all optical satellite mega-constellations. 
By enhancing visibility time and shifting the weather-related bottleneck to a more manageable, near-Earth platform, HAGS effectively navigates the complexities of weather uncertainty and transitions to an appealing buffer-based model. 
% Contribution
Within this paper, we conducted a comprehensive, simulation-based evaluation to quantify the performance advantages of HAGS over traditional ground stations (GS). 
% Findings
Our findings demonstrate that HAGS can significantly reduce the reliance on GS, under certain weather conditions. 
% Equivalency
Furthermore, we established an equivalency analysis between HAGS and GS, providing insights for strategic decision-making in satellite network design. 
% Closure
By identifying key challenges within the field of HAGS, this work catalyzes advancements toward the realization of all optical mega-constellations. 

\section*{Acknowledgments}
This work has been supported by the National Research Council Canada's (NRC) High Throughput Secure Networks program within the Optical Satellite Communications Consortium Canada (OSC) framework, Mitacs, the EU's H2020 R\&D program under the Marie Skłodowska-Curie grant agreement No 101008233 (MISSION project) and the French National Research Agency (ANR) projects ANR-22-CE25-0014-01.

\bibliography{references}
\bibliographystyle{IEEEtran}
\end{document}